\RequirePackage{fix-cm}
\documentclass[smallextended]{svjour3}       % onecolumn (second format)
\smartqed  % flush right qed marks, e.g. at end of proof
\usepackage{graphicx}
\usepackage{amsmath}
\usepackage{amssymb}
\newcommand{\ket}[1]{ | #1 \rangle}

\begin{document}

\title{Constituent-quark model with pionic contributions: electromagnetic
$N\rightarrow\Delta$ transition%\thanks{Grants or other notes
}

\titlerunning{Electromagnetic
$N\rightarrow\Delta$ transition}        % if too long for running head

\author{Ju-Hyun Jung         \and
        Wolfgang Schweiger   \and
        Elmar P. Biernat
}


\institute{J.-H. Jung \and W. Schweiger \at
              Institut f\"ur Physik, FB Theoretische Physik, Universit\"at Graz, Austria\\
               \email{ju.jung@uni-graz.at}\\
              \email{wolfgang.schweiger@uni-graz.at}         %  \\
\and E.P. Biernat \at
Centro de F\' isica Te\' orica de Part\' iculas, Instituto Superior T\'ecnico, Lisboa, Portugal\\
\email{elmar.biernat@tecnico.ulisboa.pt}
}

\date{Received: date / Accepted: date}

\maketitle

\begin{abstract}
We report on ongoing work to determine the pion-cloud contribution to the electromagnetic
$N\rightarrow\Delta$ transition form factors. The starting point is an $SU(6)$ spin-flavor symmetric constituent-quark model with instantaneous confinement that is augmented by dynamical pions which couple directly to the quarks. This system is treated in a relativistically invariant way within
the framework of point-form quantum mechanics using a multichannel formulation. The first step is to determine the electromagnetic form factors of the bare particles that consist only of three quarks. These form factors are basic ingredients for calculating the pion-cloud contributions. Already without the pion cloud, electromagnetic nucleon and $N\rightarrow \Delta$ transition form factors compare reasonably well with the data. By inclusion of the pion-cloud contribution coming from the $\pi$-$N$ intermediate state the reproduction of the data is further improved.
\end{abstract}

\section{Introduction and Formalism\label{sec:Introduction}}
Electroexcitation of the $\Delta$ resonance in electron-nucleon scattering provides important information on the structure of the $\Delta$. Although the $\Delta$ resonance was discovered several decades ago, precise experimental data became available only recently~\cite{Aznauryan:2009mx,Blomberg:2015zma}.
Many model calculations and also lattice simulations
predicted electromagnetic $N\rightarrow\Delta$ transition form
factors, indicating that the pion cloud of the nucleon and the $\Delta$ may play a substantial
role, not only in the sub-leading form factors $G_{E}^{*}$ and $G_{C}^{*}$,
but also in the leading form factor $G_{M}^{*}$~\cite{Ramalho:2008dp,Ramalho:2009df,Alexandrou:2004xn,Ledwig:2008es,Sanchis-Alepuz:2017mir}.

The electromagnetic $N\rightarrow\Delta$ transition form factors encode the structure of the $\gamma^\ast N\Delta$ vertex and show up in the covariant decomposition of the $N\rightarrow \Delta$ transition current. A common choice for the covariant decomposition of this current, involving the form factors $g_{M}$, $g_{E}$ and $g_{C}$, is given by~\cite{Pascalutsa:2006up}:
\begin{eqnarray}\label{eq:1}
J^{\mu}_{N\rightarrow\Delta}(p',\sigma';p, \sigma)  & = & i\sqrt{\frac{2}{3}}\frac{3e\left(m_{\Delta}+m_{N}\right)}{2m_{N}\left[\left(m_{\Delta}+m_{N}\right)^{2}
+Q^{2}\right]}\bar{u}_{\beta}\left(p',\sigma'\right) \nonumber \\
 &  &\times \Big\{ g_{M}\left(Q^{2}\right)\epsilon^{\beta\mu\rho\sigma}p'_{\rho}q_{\sigma} \nonumber \\
 &  & \phantom{\times} +g_{E}\left(Q^{2}\right)\left(q^{\beta}p'{}^{\mu}-q\cdot p'g^{\beta\mu}\right)i\gamma_{5} \nonumber \\
 &  &\phantom{\times} + g_{C}\left(Q^{2}\right)\left(q^{\beta}q^{\mu}-q^{2}g^{\beta\mu}\right)i\gamma_{5}\Big\} u\left(p,\sigma\right)\, .
\end{eqnarray}
Here $\bar{u}_{\beta}\left(p',\sigma'\right)$ denotes  the Rarita-Schwinger spinor of the spin-3/2 $\Delta$ and $u\left(p, \sigma\right)$ the Dirac spinor of the spin-1/2 nucleon. These spinors are normalized according to $\bar{u}_{\beta}\left(p,\sigma'\right)u^{\beta}\left(p,\sigma\right)=-2m_{\Delta}\delta_{\sigma\sigma'}$,
$\bar{u}\left(p,\sigma'\right)u\left(p,\sigma\right)=2m_{N}\delta_{\sigma\sigma'}$. Like the nucleon spinor, the Rarita-Schwinger spinor satisfies a Dirac equation, i.e. $p_{\nu}' \gamma^{\nu} {u}_{\beta}\left(p',\sigma^\prime\right)=
m_{\Delta} {u}_{\beta}(p',\sigma^\prime)$
and, in addition, the transversality condition $p^{\prime\beta} {u}_{\beta}(p',\sigma')=0=\gamma^{\beta}{u}_{\beta}\left(p',\sigma^\prime\right)$ as well as current
conservation $q_{\mu}J^{\mu}=0$, where $q=\left(p'-p\right)$ and
$q^{2}=-Q^{2}$.

For a proper relativistic description of the $N\rightarrow\Delta$ transition form factors
we make use of point-form relativistic quantum mechanics in connection with
the Bakamjian-Thomas construction. Like in previous work~\cite{Biernat:2009my,GomezRocha:2012zd,Biernat:2014dea} we use this framework to determine the one-photon-exchange amplitude for $e^- p\rightarrow e^- \Delta^+$ scattering. From this scattering amplitude we extract the electromagnetic $p\rightarrow \Delta^+$ transition current and determine the form factors by means of a covariant analysis. 
Thereby both, the nucleon and the Delta are assumed to consist of a $3q$ and a $3q$+$\pi$ component and, in addition to the dynamics of electron and quarks, the dynamics of the photon and the pion are fully taken into account. This is accomplished by means of a multichannel formulation that comprises all states which can occur during the scattering process (i.e. $|3q, e \rangle$, $|3q, \pi, e \rangle$, $|3q, e, \gamma \rangle$, $|3q, \pi, e, \gamma \rangle$). What one then needs, in principle, are scattering solutions of
\begin{equation}\label{EVequation}
\left(\begin{array}{cccc}
\hat{M}_{3qe}^{\mathrm{conf}} & \hat{K}_\pi & \hat{K}_\gamma & \hat{K}_{\pi\gamma}
\\
\hat{K}_\pi^\dagger & \hat{M}_{3q \pi e} ^{\mathrm{conf}}& \hat{K}_{\pi\gamma}^\prime &
\hat{K}_\gamma \\
\hat{K}_\gamma^\dagger & \hat{K}_{\pi\gamma}^{\prime\dag} & \hat{M}_{3q e \gamma} ^{\mathrm{conf}}&
\hat{K}_\pi \\
\hat{K}_{\pi\gamma}^\dag & \hat{K}_\gamma^\dagger & \hat{K}_\pi^\dagger &
\hat{M}_{3q \pi e \gamma}^{\mathrm{conf}}
\end{array}\right)
\left(\begin{array}{l}
\ket{\psi_{3q e}} \\ \ket{\psi_{3q \pi e}} \\
\ket{\psi_{3q e \gamma}} \\ \ket{\psi_{3q \pi e \gamma}}
\end{array}\right)
=
\sqrt{s} \left(\begin{array}{l}
\ket{\psi_{3q e}} \\ \ket{\psi_{3q \pi e}} \\
\ket{\psi_{3q e \gamma}} \\ \ket{\psi_{3q \pi e \gamma}}
\end{array}\right)
\end{equation}
which evolve from an asymptotic electron-nucleon in-state $\ket{e N}$ with invariant mass $\sqrt{s}$ into an asymptotic electron-Delta out-state $\ket{e \Delta}$. The diagonal entries of this matrix mass operator contain, in addition to the relativistic kinetic energies of the particles in the particular channel, an instantaneous confinement potential between the quarks. The off-diagonal entries are vertex operators which describe the transition between the channels. In a velocity-state representation these vertex operators are directly related to  usual quantum-field theoretical interaction-Lagrangean densities~\cite{Biernat:2010tp}. The 4-vertices $\hat{K}_{\pi\gamma}$ and $\hat{{K}}_{\pi\gamma}^{\prime}$ show up only for pseudovector pion-quark coupling. These vertices are neglected in the present form of the model, but obviously have to be included in an improved version.

\begin{figure}[tb]
\center{\includegraphics[width=0.35\textwidth]{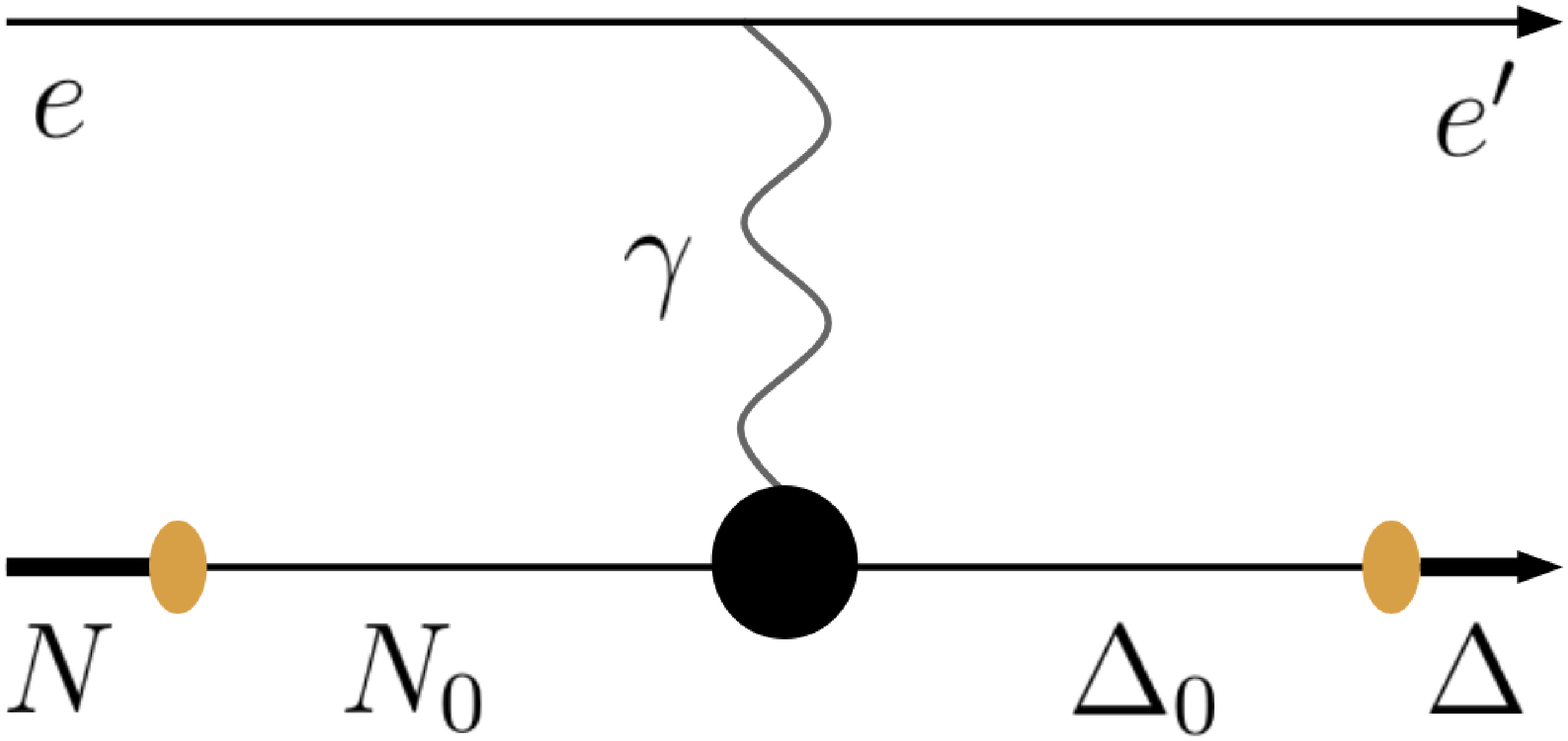}\\ \medskip
\includegraphics[width=0.35\textwidth]{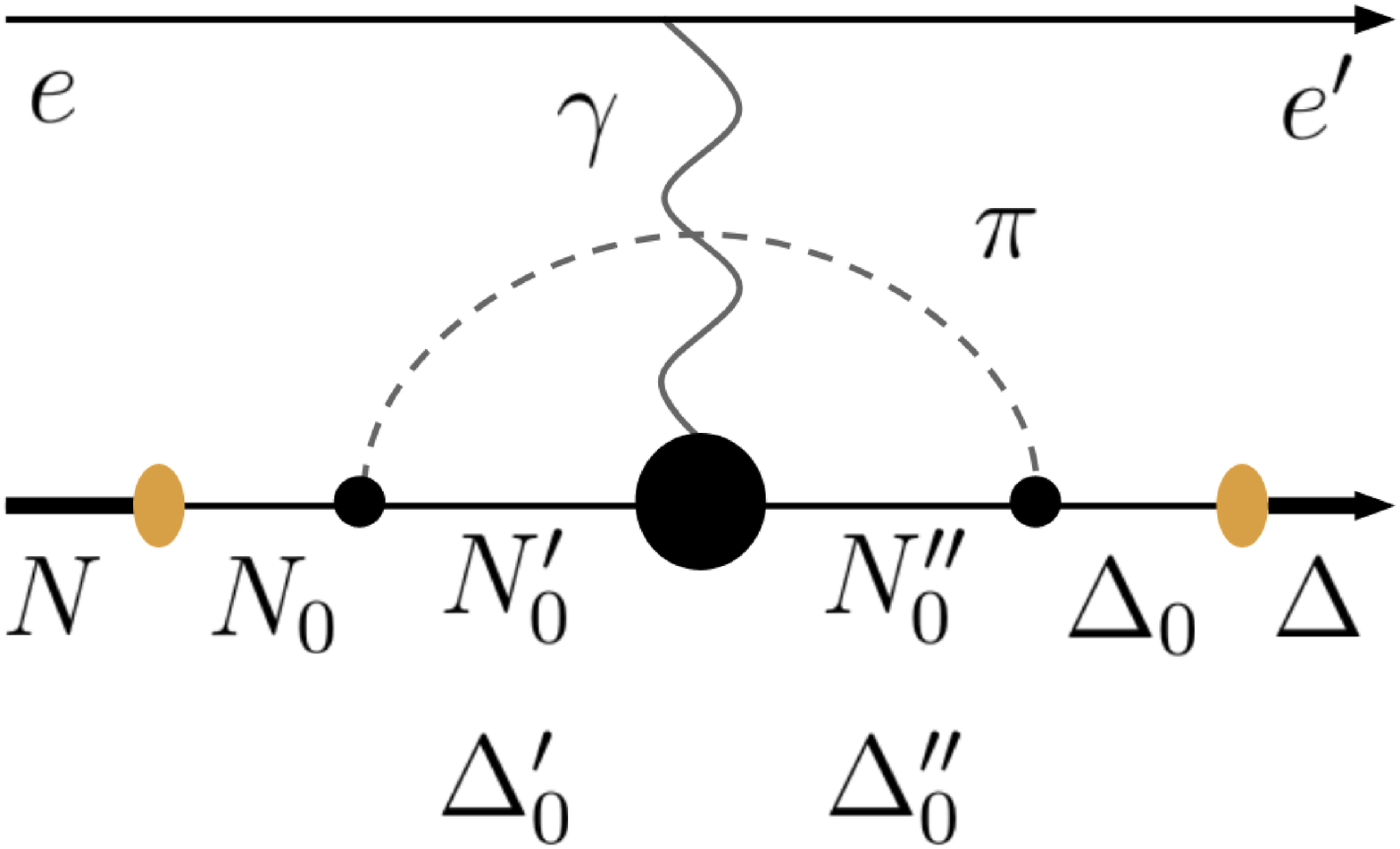} \hspace{1.0cm} \includegraphics[width=0.35\textwidth]{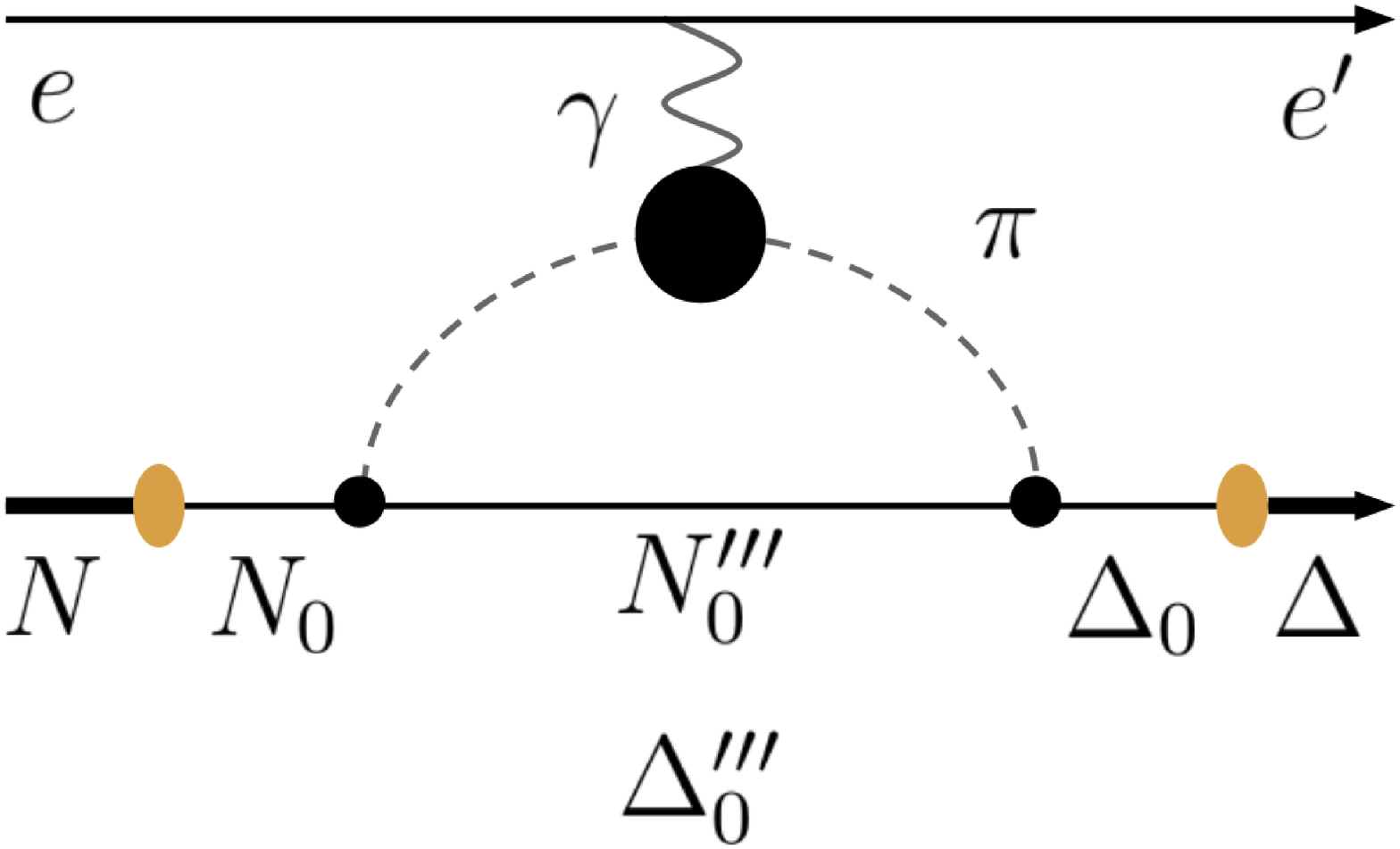}
}
\vspace{-0.2cm}
\caption{The three graphs contributing to electroexcitation of the $\Delta$ resonance in the presence of a pion cloud. The big blobs represent electromagnetic (transition) form factors involving the bare nucleon $N_0$ and the bare Delta $\Delta_0$. The small black blobs represent strong form factors at the $\pi N_0 N_0$, $\pi N_0 \Delta_0$ and $\pi\Delta_0 \Delta_0$ vertices. All these form factors are determined by the valence-quark wave functions of the bare baryons. A vertex form factor, calculated within a constituent-quark model~\cite{Biernat:2009my} and the same approach as used here, is also assumed at the pion-photon vertex.}
\label{fig:1}
\vspace{-0.3cm}
\end{figure}

At this point it is convenient to reduce Eq.~(\ref{EVequation}) to an eigenvalue problem for $\ket{\psi_{3q e}} $ by means of a Feshbach reduction:
\begin{equation}\label{eq:Mphys}
\left[\hat{M}_{3qe}^{\mathrm{conf}} +\hat{K}_\pi(\sqrt{s}-\hat{M}_{3q\pi e}^{\mathrm{conf}} )^{-1} \hat{K}_\pi^\dag + \hat{V}_{1\gamma}^{\mathrm{opt}}(\sqrt{s})\right] \ket{\psi_{3q e}} = \sqrt{s} \, \ket{\psi_{3q e}} \, .
\end{equation}
Here $\hat{V}_{1\gamma}^{\mathrm{opt}}(\sqrt{s})$ is the 1$\gamma$-exchange optical potential. The invariant 1$\gamma$-exchange amplitude for electroproduction of the Delta is now obtained by sandwiching $\hat{V}_{1\gamma}^{\mathrm{opt}}(\sqrt{s})$ between (the valence component of) physical electron-nucleon  $\ket{eN}$ and electron-Delta $\ket{e\Delta}$ states , i.e. eigenstates of $[  \hat{M}_{3qe}^{\mathrm{conf}} +\hat{K}_\pi (\sqrt{s}-\hat{M}_{3q\pi e}^{\mathrm{conf}} )^{-1} \hat{K}_\pi^\dag ]$. The crucial point is now to observe that, due to instantaneous confinement, propagating intermediate states do not contain free quarks, they rather contain bare nucleons $N_0$ or bare Deltas $\Delta_0$. The bare particles are eigenstates of the pure confinement problem. This allows us to rewrite the scattering amplitude in terms of pure hadronic degrees of freedom with the quark substructure being hidden in strong and electromagnetic vertex form factors of the bare baryons. This is graphically represented in Fig.~\ref{fig:1}.

In order to calculate the graphs shown in Fig.~\ref{fig:1} we obviously have to know the structure of the strong and electromagnetic vertices for bare baryons and also the masses of the bare nucleon and the bare Delta. For scalar, isoscalar confinement these masses and also the three-quark wave functions are the same due to $SU(6)$ spin-flavor symmetry. Instead of choosing a particular confining interaction we therefore rather parameterize the three-quark wave function of $N_0$ and $\Delta_0$ by means of a Gaussian. Knowing further the bare nucleon and Delta mass $m_{N_0}=m_{\Delta_0}=:m_0$, the (pseudovector) pion-quark coupling $f_{\pi q q}$ and the constituent-quark masses $m_u=m_d=:m_q$, one can first calculate the strong couplings and form factors at the $\pi N_0 N_0$, $\pi N_0 \Delta_0$ and $\pi N_0 \Delta_0$ vertices and in the sequel the renormalization effect of pion loops on the nucleon and Delta mass. One thus has a four-parameter model which provides a microscopic description of a coupled system of nucleons, pions and Deltas. With the constituent-quark mass $m_q=0.263$~GeV taken from the literature~\cite{Pasquini:2007iz}, the remaining three parameters can be adapted in such a way that the solution of a mass-eigenvalue problem analogous to Eq.~(\ref{EVequation}) (just without electron and photon) gives the physical nucleon and Delta masses. A more detailed account of how strong form factors and couplings are calculated and the model parameters are fixed can be found in Ref.~\cite{Jung:2017cpy}. The values used here are $\alpha=2.645$~GeV$^{-1}$ for the range of the Gaussian, $f_{\pi q q}=0.678$ for the pseudovector pion-quark coupling and $m_0=1.67$~GeV for the bare nucleon and Delta mass.\footnote{These values differ slightly from those given in Ref.~\cite{Jung:2017cpy}, since the numerics in this paper was still afflicted by an error in the computer program.}

The pion-baryon couplings and form factors are thus already determined after having fixed the parameters of the model in the way just sketched. What is still necessary to calculate the leading order electroproduction amplitude as depicted in Fig.~\ref{fig:1}, are the electromagnetic (transition) form factors of the bare baryons. We follow the same strategy as outlined above, but neglect the pion, to end up with the one-photon exchange amplitude $\mathcal{M}_{1\gamma}^{e B_0\rightarrow e B_0^\prime}$ for $e B_0\rightarrow e B_0^\prime$ scattering. As one would expect, this amplitude can be written as (covariant) photon  propagator times electron current contracted with the baryonic current, $\mathcal{M}_{1\gamma}^{e B_0\rightarrow e B_0^\prime}\propto j_{e\mu} I^\mu_{B_0\rightarrow B_0^\prime}/Q^2$. This allows to extract a microscopic expression for the baryonic current $I^\mu_{B_0\rightarrow B_0^\prime}$, which turns out to be an integral over the three-quark wave functions of incoming and outgoing baryons, multiplied with the electromagnetic quark current and some Wigner-rotation factors~\cite{Biernat:2009my}. 

By means of a general covariant decomposition of $I^\mu_{B_0\rightarrow B_0^\prime}$ one would then be able to identify the electromagnetic (transition) form factors of the bare baryons. But here a problem shows up. $I^\mu_{N_0\rightarrow \Delta_0}$, e.g., is expected to have the structure given in Eq.~(\ref{eq:1}). Numerical studies, however, reveal that one needs additional covariants for a complete covariant decomposition of $I^\mu_{N_0\rightarrow \Delta_0}$, which involve an electron momentum. In addition, the form factors in front of the covariants do not only depend on the square of the transferred four momentum $q^2=-Q^2$, but also on the invariant mass $\sqrt{s}$ of the electron-baryon system. It is an unwanted feature, but does not spoil the relativistic invariance of $\mathcal{M}_{1\gamma}^{e N_0\rightarrow e \Delta_0^\prime}$.  A similar observation has already been made in Refs.~\cite{Biernat:2009my} and \cite{Biernat:2014dea} when calculating electromagnetic $\pi$ and $\rho$ form factors within a constituent-quark model using the same approach as here. There it turned out that the non-physical, spurious contributions to the electromagnetic current vanish, or become at least minimal for large invariant mass of the electron-hadron system. Sensible results for the form factors were obtained in the limit $s\rightarrow \infty$. This limit corresponds to the kinematical situation that the subprocess $\gamma^\ast H\rightarrow H$ is considered in the infinite-momentum frame and momentum is transferred in transverse direction. 

Before we continue, we want to make a few remarks about the observation that our microscopic current exhibits spurious contributions. An analogous situation occurs in the covariant light-front approach presented in Ref.~\cite{Carbonell:1998rj}, where spurious contributions to the current are connected with the four vector that describes the orientation of the light front. Actually it turned out that the results for the physical $\pi$ and $\rho$ form factors in Refs.~\cite{Biernat:2009my} and \cite{Biernat:2014dea} are the same as corresponding results obtained within the covariant light-front approach. One should also mention that most models for electromagnetic bound-state currents are formulated in a particular reference frame, usually the Breit frame, and the frame dependence of the resulting form factors is kept under the carpet.  The advantage of our approach is, that we have some control on the frame dependence and extract the form factors in a frame, namely the infinite momentum frame, in which this dependence vanishes. In our case, the reason that the electromagnetic hadron current exhibits some dependence on the momentum of the scattered electron is most likely that the Bakamjian-Thomas construction, used to implement interactions without spoiling relativistic invariance, causes problems with cluster separability. These problems can be cured by appropriate unitary transformations which are formally known~\cite{Sokolov:1977ym}, but technically hard to realize. Work in this direction is in progress. 

Here we follow the same strategy as in Refs.~\cite{Biernat:2009my} and \cite{Biernat:2014dea} and go to the infinite-momentum frame to extract the electromagnetic form factors from our microscopic expressions for the currents, $I^\mu_{B_0\rightarrow B_0^\prime}$. Concentrating first on the $N\rightarrow \Delta$ transition we observe that the physical current, as given in Eq.~(\ref{eq:1}), has only four different spin-matrix elements in leading order in $k=\sqrt{s}/2$. These are
\begin{eqnarray}\label{eq:J13}
J_{N\rightarrow\Delta}^{0}\left(\frac{3}{2},\frac{1}{2}\right) \hspace{0.2cm}
& = & \chi_Q\left\{ \sqrt{3}\left[g_{M}\left(m_{N}+m_{\Delta}\right) \right. 
\!+\! \left.g_{E}\left(m_{N}-m_{\Delta}\right)\right]kQ+\mathcal{\mathcal{O}}\left(\frac{1}{k}\right)\right\},\nonumber \\ \\
J_{N\rightarrow\Delta}^{0}\left(\frac{1}{2},\frac{1}{2}\right) \hspace{0.2cm}
& = &  \chi_Q\left\{ \left[-g_{M}+g_{E}\right]kQ^{2} +\frac{2g_{C}\left(m_{N}-m_{\Delta}\right)kQ^{2}}{m_{\Delta}}+\mathcal{O}\left(\frac{1}{k}\right)\right\},\nonumber \\ \\
J_{N\rightarrow\Delta}^{0}\left(-\frac{1}{2},\frac{1}{2}\right) 
& = & \chi_Q\bigg\{ \left[g_{M}\left(m_{N}+m_{\Delta}\right)-g_{E}\left(m_{N}-m_{\Delta}\right)\right]kQ\nonumber\\ & & \hspace{0.8cm}\left.
+\frac{2g_{C}kQ^{3}}{m_{\Delta}}+\mathcal{O}\left(\frac{1}{k}\right)\right\},\\ \label{eq:J1m3}
J_{N\rightarrow\Delta}^{0}\left(-\frac{3}{2},\frac{1}{2}\right) 
& = &  \chi_Q\left\{ \sqrt{3}\left[-g_{M}-g_{E}\right]k Q^{2}+\mathcal{O}\left(\frac{1}{k}\right)\right\} \, ,
\end{eqnarray}
where $\chi_Q=(m_{\Delta}+m_{N})/(2 m_N [(m_{\Delta}+m_{N})^{2}+Q^{2}])$. The remaining spin-matrix elements of the current are either related by parity, they are identical, or they vanish. Since there are four spin-matrix elements, but only three form factors, these spin-matrix elements cannot be independent, but must be linearly related. This relation is, what one calls the \lq\lq angular condition\rq\rq~\cite{Carbonell:1998rj} and has the form:
\begin{eqnarray}\label{eq:angcond}
&&J_{N_{\frac{1}{2}}\rightarrow\Delta_{\frac{3}{2}}}^{0}\left(M_{N}^{2}-M_{N}M_{\Delta}+Q^{2}\right)Q+\sqrt{3}J_{N_{\frac{1}{2}}\rightarrow\Delta_{\frac{1}{2}}}^{0}M_{\Delta}Q^{2} \nonumber\\
&&\hspace{0.5cm}+\sqrt{3}J_{N_{\frac{1}{2}}\rightarrow\Delta_{-\frac{1}{2}}}^{0}\left(-M_{N}M_{\Delta}+M_{\Delta}^{2}\right)Q \nonumber\\
&&\hspace{0.5cm}+J_{N_{\frac{1}{2}}\rightarrow\Delta_{-\frac{3}{2}}}^{0}\left(\left(M_{N}-M_{\Delta}\right)^{2}\left(M_{N}+M_{\Delta}\right)+M_{N}Q^{2}\right) =  0\,, 
\end{eqnarray}
where $J_{N_{\sigma}\rightarrow\Delta_{\sigma'}}^{0}:=\lim_{k\rightarrow\infty} J_{N\rightarrow\Delta}^{0} (\sigma',\sigma)$.
What we observe is that, due to the unphysical contributions which we pick up in our approach, neither the microscopic model for the bare transition current $I_{N_0\rightarrow \Delta_0}^\mu$, nor the one including the pion cloud $I_{N\rightarrow \Delta}^\mu$ satisfy this angular condition. A way out would be to make a complete covariant decomposition of the microscopic current, involving additional, unphysical covariants. In this case the right-hand side of the angular condition would become a combination of unphysical contributions and after separating them, one would get a model for the current with the desired properties. This strategy has been pursued in Ref~\cite{Biernat:2014dea} for the $\rho$. As a first attempt, we have rather tried to extract the form factors from the different possible choices of three spin-matrix elements out of the four given in Eqs.~(\ref{eq:J13})-(\ref{eq:J1m3}). Most reasonable results for the form factors are obtained with the combination $I_{N\rightarrow \Delta}^0(\frac{3}{2},\frac{1}{2})$,
$I_{N\rightarrow \Delta}^0(-\frac{1}{2},\frac{1}{2})$ and $I_{N\rightarrow \Delta}^0(-\frac{3}{2},\frac{1}{2})$. A similar strategy was adopted in Ref.~\cite{Cardarelli:1995dc} to calculate $N\rightarrow \Delta$ transition form factors within a front-form approach.

Similar problems with the angular condition are also expected to show up when calculating electromagnetic $\Delta$ form factors. For the nucleon, however, only two independent current matrix elements come with $\mathcal{O}(k)$ in the infinite-momentum frame, allowing for a unambiguous extraction of the electromagnetic nucleon form factors~\cite{Kupelwieser:2015ioa}.

\section{Results and discussion}
In the present calculation only the $N_0 \pi$ state is taken into account in the pion loop. Therefore we do not need the electromagnetic form factors of the (bare) $\Delta$ at this stage. Having fixed the parameters of the model as described above, we already know the strong $\pi N_0 N_0$ and $\pi N_0 \Delta_0$ couplings and form factors. In a next step we calculate the electromagnetic $N_0$ and $N_0\rightarrow \Delta_0$ form factors.  These are then used to determine the one-photon-exchange amplitude as given in Fig.~\ref{fig:1}. From this amplitude we extract the microscopic transition current $I^\mu_{N\rightarrow \Delta}$ for the physical (dressed) nucleon and Delta and, in the sequel, the electromagnetic transition form factors, taking the spin-matrix elements $I_{N\rightarrow \Delta}^0(\frac{3}{2},\frac{1}{2})$, $I_{N\rightarrow \Delta}^0(-\frac{1}{2},\frac{1}{2})$ and $I_{N\rightarrow \Delta}^0(-\frac{3}{2},\frac{1}{2})$ (see discussion above). 

The electromagnetic form factors $g_M$, $g_E$ and $g_C$, introduced in Eq.~(\ref{eq:1}), relate to the more conventional magnetic dipole $G_{M}^{*}$, electric quadrupole $G_{E}^{*}$ and Coulomb quadrupole $G_{C}^{*}$ form factors of Jones and Scadron~\cite{Jones:1972ky} as follows:
\begin{eqnarray}
G_{M}^{*} & = & g_{M}+G_{E}^{*}\, , \\
G_{E}^{*} & = & \frac{1}{\left(M_{\Delta}+M_{N}\right)^{2}+Q^{2}}\left[\frac{1}{2}\left(-M_{\Delta}^{2}+M_{N}^{2}+Q^{2}\right)g_{E}+Q^{2}g_{C}\right]\, , \\
G_{C}^{*} & = & \frac{1}{\left(M_{\Delta}+M_{N}\right)^{2}+Q^{2}}\left[\left(-M_{\Delta}^{2}+M_{N}^{2}+Q^{2}\right)g_{C}-2M_{\Delta}^{2}g_{E}\right]\, .
\end{eqnarray}
In the following we will present our results in terms of these form factors.
\begin{figure}\label{fig:2}
\center{\includegraphics[width=0.50\textwidth]{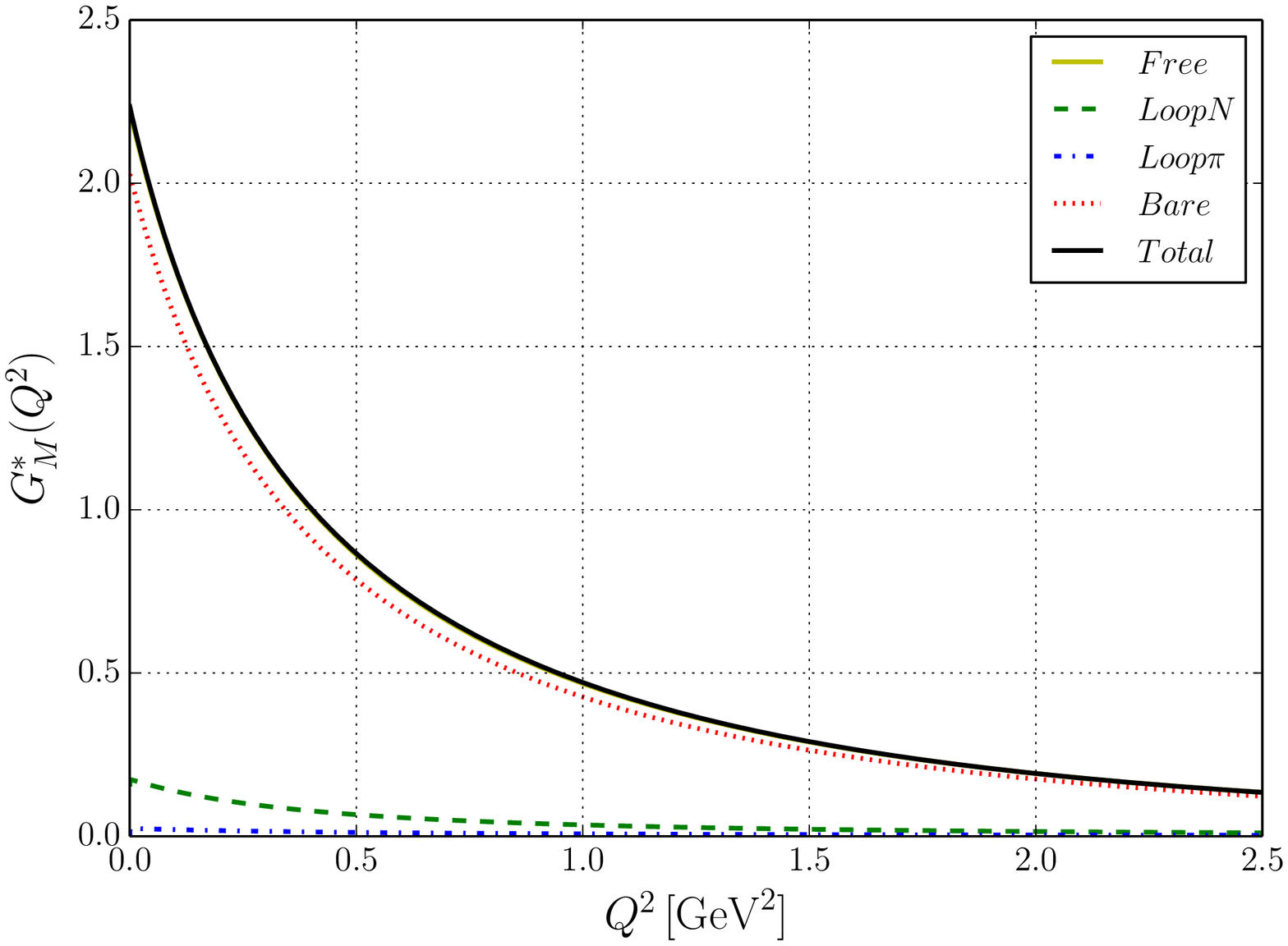}\includegraphics[width=0.50\textwidth]{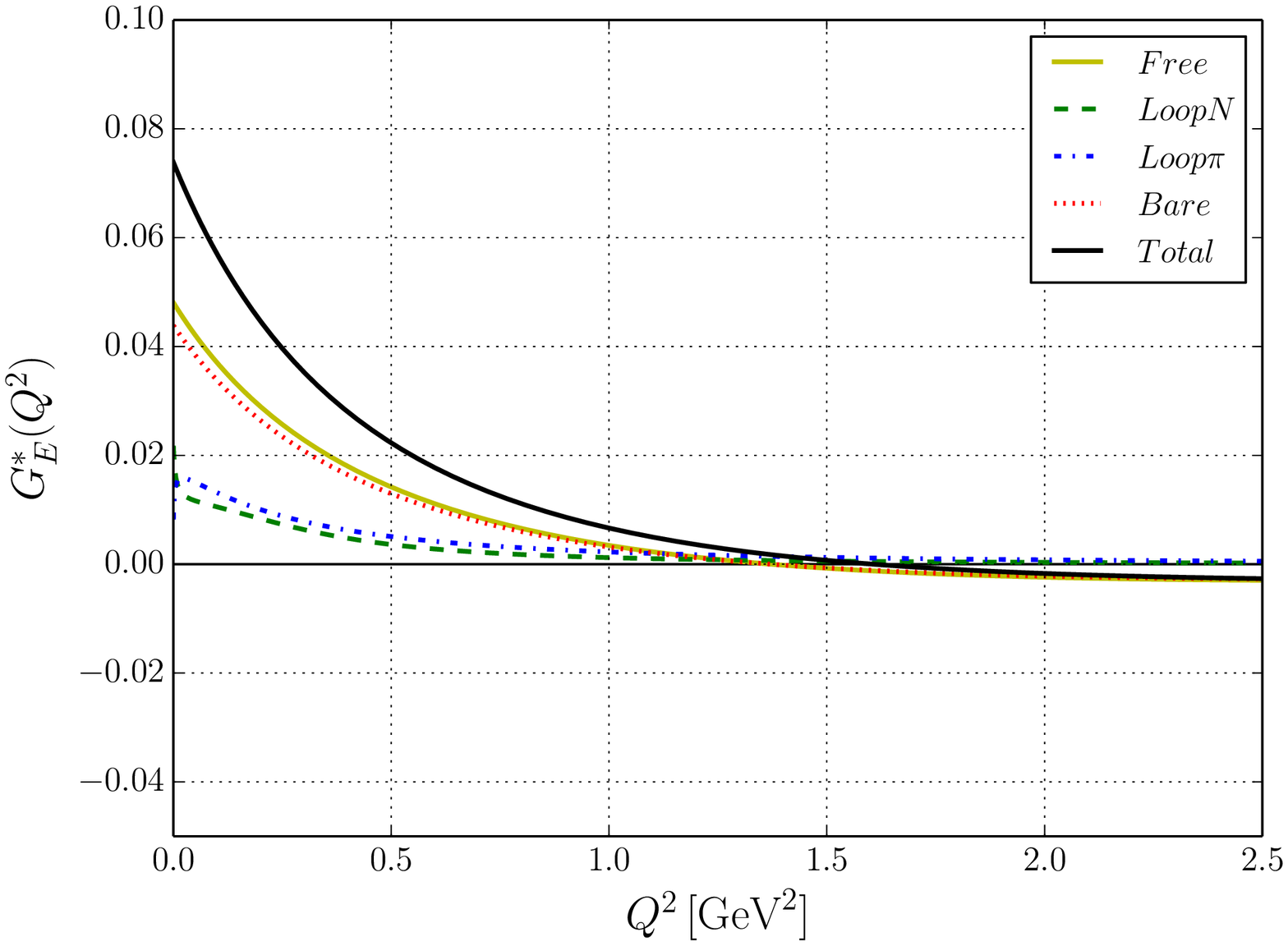}

\includegraphics[width=0.50\textwidth]{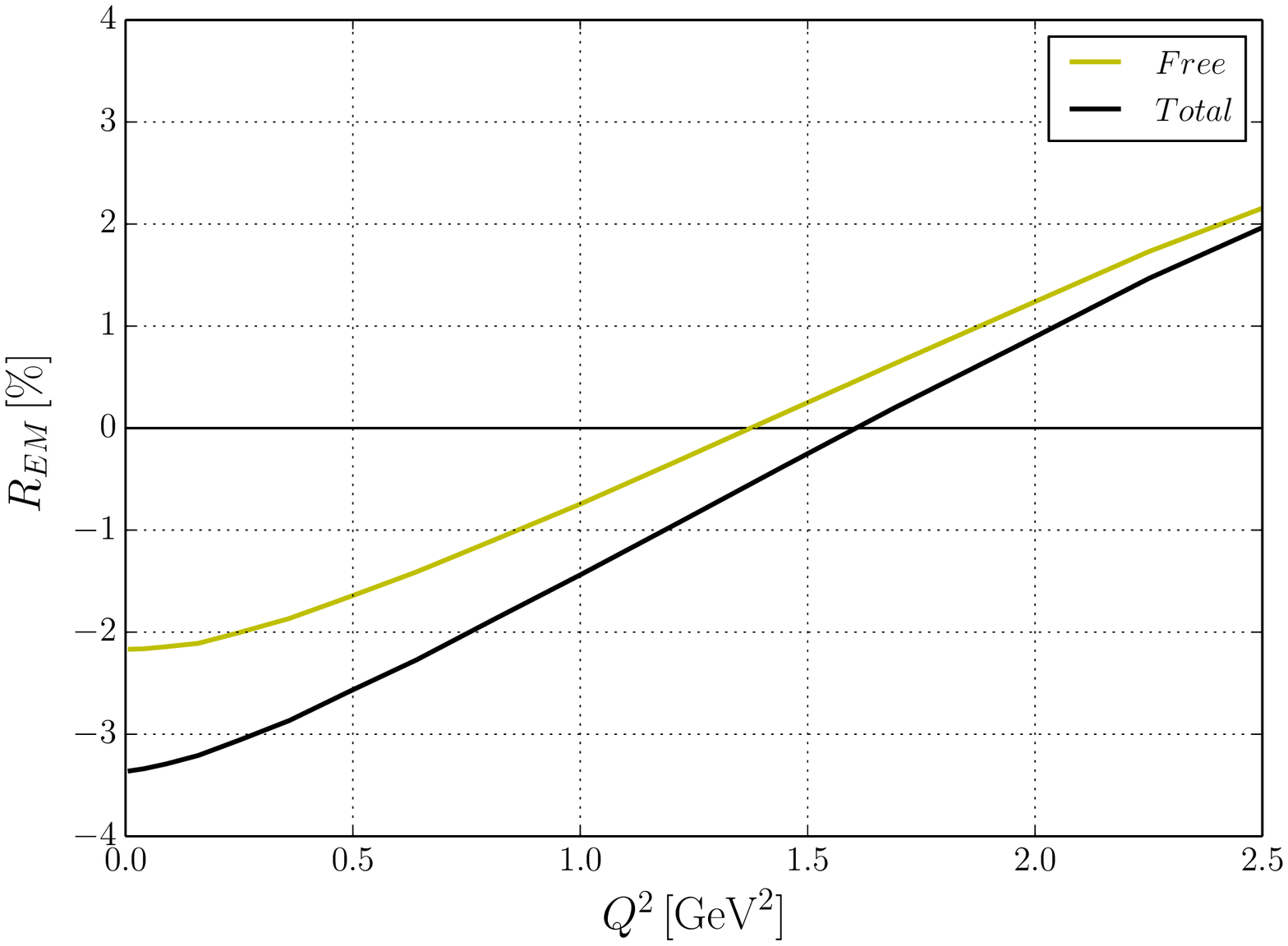}
}
\caption{The upper left and the upper right panel show our model predictions for the magnetic dipole $G_{M}^{*}$ and the electric quadrupole $G_{E}^{*}$ form factors, respectively. The lower panel depicts the $R_{EM}$ ratio in percent. The label \lq\lq Free\rq\rq\  refers to a calculation without pion-loop contributions (bare and physical particles are the same). The labels \lq\lq Bare\rq\rq, \lq\lq LoopN\rq\rq\ and  \lq\lq Loop$\pi$\rq\rq\  refer to the contributions of the first, the second and the third graph of Fig.~\ref{fig:1}, respectively. \lq\lq Total\rq\rq\ means the sum of all three graphs.}
\label{fig:2}
\end{figure}

Fig.~\ref{fig:2} shows the magnetic dipole form factor $G_{M}^{*}$, the electric quadrupole form factor $G_{E}^{*}$, and their ratio $R_{EM}$. Pion-cloud effects, seemingly, do not play a role for $G_{M}^{*}$. This does not mean that they are negligible as compared to the bare-baryon contribution (first graph in Fig.~\ref{fig:1}). The bare contribution is nothing else than the result without pion cloud (pure three-quark model) multiplied with  the probabilities to find the bare baryons in the physical (dressed) ones. The resulting reduction is then again compensated by the pion cloud. Pion-cloud effects, however, become more visible in the small form factors $G_{E}^{*}$ and $G_{C}^{*}$. Here we only show predictions for $G_{E}^{*}$ and the ratio
\begin{eqnarray}
R_{EM} & := & -\frac{G_{E}^{*}\left(Q^{2}\right)}{G_{M}^{*}\left(Q^{2}\right)}\, .
\end{eqnarray}
For these quantities pion-cloud effects seem to be significant, at least for $Q^2\lesssim 1$~GeV$^2$, with both contributions, the one in which the photon couples to the pion and the one in which it couples to the nucleon,  being of approximately the same importance. Our results compare with the outcome of other theoretical predictions~\cite{Ramalho:2008dp,Sanchis-Alepuz:2017mir,Cardarelli:1995dc}. For $Q^2\gtrsim 0.5$~GeV$^2$ our predictions for $G_{M}^{*}$ agree well with the data, for vanishing $Q^2$, however, we underestimate the data by about $25\%$. This is also reflected in $R_{EM}$.  For For $Q^2\lesssim 0.5$~GeV$^2$ we get a somewhat larger modulus for this ratio than measured in experiment. One should, however, keep in mind that our calculation is still not complete and additional contributions at small $Q^2$ are expected to come from  $\pi \Delta_0$ intermediate states. It is the topic of ongoing work to find out, whether such contributions could improve the agreement with data, or whether further improvements of the model, like a more sophisticated wave function for the $\Delta$, including, e.g., a $d$-wave contribution as in Ref.~\cite{Ramalho:2008dp}, will be necessary.

\begin{acknowledgements}
J.-H. Jung acknowledges the support of the Fonds zur F\"orderung der wissenschaftlichen Forschung in \"Osterreich (Grant No. FWF DK W1203-N16). He furthermore wants to thank Prof. T. Pe\~na for giving him the opportunity to stay at the Centro de F\'isica Te\'orica de Part\'iculas, IST Lisboa, where part of this work was done. E. P. Biernat acknowledges the support of the Funda\c c\~ao para a Ci\^encia e a 
Tecnologia (FCT) under Grant No. SFRH/BPD/100578/2014.
\end{acknowledgements}


\end{document}